\documentclass[twocolumn,showpacs,preprintnumbers,amsmath,amssymb,prl]{revtex4}

\usepackage{graphicx}
\usepackage{dcolumn}
\usepackage{bm}

\begin{document}
\title{Intrinsic aging and effective viscosity in the slow dynamics of a soft glass with tunable
elasticity}
\author{Laurence Ramos and Luca Cipelletti}
\affiliation{Groupe de Dynamique des Phases Condens\'{e}es (UMR
CNRS-UM2 5581), CC26, Universit\'{e} Montpellier 2, 34095
Montpellier Cedex 5, France}

\email{ramos@gdpc.univ-montp2.fr}
\date{\today}

\begin{abstract}
We investigate by rheology and light scattering the influence of
the elastic modulus, $G_0$, on the slow dynamics and the aging of
a soft glass. We show that the slow dynamics and the aging can be
entirely described by the evolution of an effective viscosity,
$\eta_{eff}$, defined as the characteristic time measured in a
stress relaxation experiment times $G_0$. At all time,
$\eta_{eff}$ is found to be independent of $G_0$, of elastic
perturbations, and of the rate at which the sample is quenched in
the glassy phase. We propose a simple model that links
$\eta_{eff}$ to the internal stress built up at the fluid-to-solid
transition.
\end{abstract}

\maketitle



A large variety of disordered soft solids exhibit fast dynamics
that are associated with overdamped elastic modes activated by
thermal energy. As a consequence, the fast dynamics of systems
such as polyelectrolyte or colloidal gels
\cite{Cloitre2003,KrallPRL1998}, concentrated emulsions
\cite{ccemulsions} and surfactant or amphiphilic copolymer phases
\cite{Onions_CRPP, Faraday} is related to the elastic modulus,
$G_0$, of the material. In addition, most of these systems exhibit
a slow and non stationary dynamics, whose origin is still under
debate. One emerging idea is that the slow dynamics be due to the
relaxation of internal stress built up at the fast transition from
a fluid state to a solid state \cite{Faraday, int_stresses, paper
Munch PRE, Estelle}. In this scenario, the internal stress would
result from a deformation of the local structure with respect to
the ideal, relaxed configuration and thus would be proportional to
the elasticity of the system. Accordingly, one may expect the slow
dynamics to be intimately connected to the elastic properties of
the material. However, experiments that directly test these ideas
are still lacking, due to the difficulty of quenching a system
into a glassy phase without perturbing the internal stress
distribution. Indeed, for most soft glasses the fluid-to-solid
transition is obtained upon cessation of a large preshear
\cite{Cloitre2000, Bonn2002,Derec2003}, which certainly influences
the initial configuration of internal stress
\cite{ViasnoffPRL2002}.

In this Letter, we investigate the slow dynamics of a disordered
soft material, for which $G_0$ can be varied over more than one
decade without changing significantly the structure, and whose
dynamics can be initialized without imposing any shear. The sample
is formed by a compact arrangement of soft and polydisperse
elastic spheres, where the transition from a fluid to a solid
state can be controlled by varying the temperature. In a previous
paper \cite{Ramos2001} we have used multispeckle dynamic light
scattering (MDLS) and linear rheology to show that this system
exhibits slow dynamics, due to the rearrangement of the spheres,
whose characteristic time increases as a power law of sample age,
$t_w$. Here, we address explicitly the question of how the
elasticity of the system influences the slow dynamics. We show
that $\tau_R$, the characteristic relaxation time of the slow
dynamics as measured in a stress relaxation experiment, depends
only on the age of the sample and its elastic modulus. Remarkably,
at all ages $\tau_R$ is found to be inversely proportional to
$G_0$, thus suggesting that the slow dynamics and the aging of all
samples can be described by the time evolution of a single
parameter, an effective viscosity defined as $\eta_{eff} =  \tau_R
G_0$. We moreover show that the evolution of $\eta_{eff}$ is
independent of perturbations of the elastic modulus and of the
rate at which the sample is quenched in the glassy phase. We
propose a simple model that relates the age-dependent effective
viscosity to the relaxation of the internal stress.


The samples are surfactant lamellar phases constituted of a
regular unidimensional stacking of bilayers that spontaneously
roll up, resulting in a dense packing of multilamellar vesicles
(MLVs) (due to their polydispersity and softness, the volume
fraction of the MLVs is one). Bilayers are composed of a mixture
of cetylpyridinium chloride (CpCl) and octanol (Oct) (weight ratio
$\rm{CpCl/Oct}=0.95$) and diluted in brine ($\rm{[NaCl]}= 0.2
\rm{M}$) at a weight fraction $\phi$, which ranges from $12 \%$ to
$16 \%$ \cite{Ligoure}. The bilayers are decorated by an
amphiphilic copolymer, Symperonics F68 by Serva
($\rm{(EO)_{76}}-\rm{(PO)_{29}}-\rm{(EO)_{76}}$, where EO is
ethylene oxide and PO is propylene oxide). The
copolymer-to-bilayer weight ratio $\alpha$ ranges between $0.2$
and $1.6$. Upon copolymer addition, a transition from a flat
lamellar phase to a MLV phase occurs. MLVs are polydisperse with a
maximum size of a few microns \cite{equilibriumonions}. The MLV
phase behaves mechanically as a gel with a frequency-independent
storage modulus, $G'$, about one order of magnitude larger than
the loss modulus, $G''$. Experimentally, we take $G_0 \approx
G'(\nu = 1 \, {\rm Hz})$. A simple model \cite{equilibriumonions}
relates $G_0$ to the MLV size distribution and the repulsive
interaction between the bilayers. Both parameters depend on $i)$
the interlamellar distance, set by $\phi$; $ii)$ the amount of
copolymer adsorbed to the bilayer, $\alpha$; $iii)$ the
hydrophobicity of the PO block of the copolymer, controlled by the
temperature, $T$. Therefore, the elastic modulus can be varied
experimentally by changing three independent parameters, $\phi$,
$\alpha$, and $T$.

\begin{figure}
\includegraphics{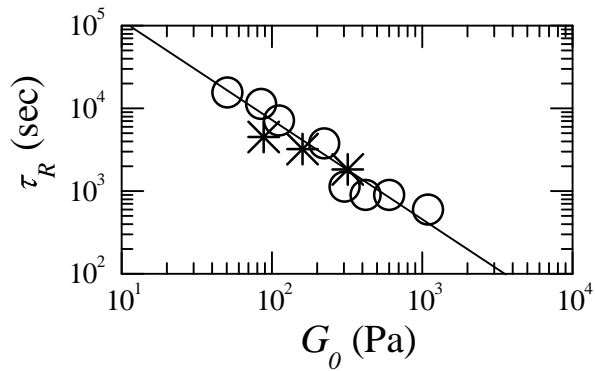}
\caption{Elastic modulus dependence of the characteristic time
measured by linear rheology for samples with $\phi=16\%$. Circles:
$T=20^\circ \rm{C}$ and $0.2 \leq \alpha \leq1.6$; Stars:
$\alpha=0.8$ and $T=11,13.9$ and $29.7^\circ \rm{C}$. The straight
line is a fit to a power law yielding an exponent $-1.1 \pm 0.1$.
All measurements are done at the same age $t_w \simeq 24000$ sec.}
\label{FIG:1}
\end{figure}

A temperature jump from $T=4^\circ \rm{C}$ to $T \geq 10^\circ
\rm{C}$ induces a fast transition from a fluid to a solid state
and initializes the dynamics of the system (we take $t_w=0$ as the
time where the sample becomes a solid, as indicated by $G'(1
\rm{Hz}) \geq \textit{G}'' (1 \rm{Hz})$). The slow dynamics is
subsequently probed by rheology measurements. The stress
relaxation that follows a step strain imposed in the linear regime
exhibits a very slow stretched exponential decay, $\sigma(\tau) =
\sigma_0 \exp[-(\tau/\tau_R)^p]$, with $p$ in the range $0.2-0.4$,
depending on sample composition \cite{Note_uncertainties}. In
fig.~\ref{FIG:1}, we show $\tau_R$, extracted from the stretched
exponential fit of $\sigma(\tau)$, as a function of the elastic
modulus $G_0$ for samples with the same age ($t_w \simeq 24000$
sec) and surfactant weight fraction ($\phi=16 \%$), but different
$G_0$. The elastic modulus was varied either by changing copolymer
content $\alpha$ at a fixed temperature ($0.2 \leq \alpha
\leq1.6$, $T=20^ \circ \rm{C}$, circles), or by varying $T$ at
fixed $\alpha$ ($11^ \circ \rm{C} \leq T \leq 30^ \circ \rm{C}$,
$\alpha=0.8$, stars). Strikingly, both sets of data collapse onto
a master curve, for which $\tau_R$ monotonically decreases as
$G_0$ increases, thus indicating that the slow dynamics is faster
for harder systems, independently of the detailed sample
composition. A power law fit to the data yields an exponent $-1.1
\pm 0.1$, suggesting that the slow dynamics may be described by
introducing an effective viscosity defined by
$\eta_{eff}=\tau_RG_0$, whose value is independent of the
composition and of the elasticity of the material over more than
one order of magnitude in $G_0$.

\begin{figure}
\includegraphics{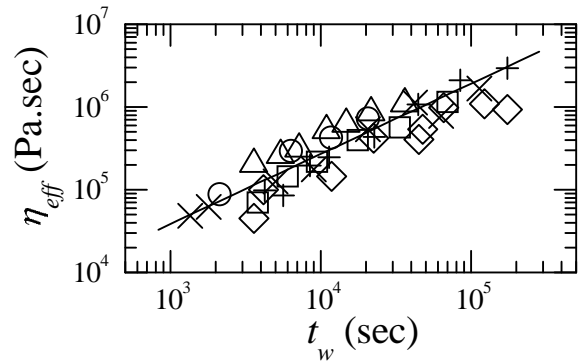}
\caption{Age dependence of the effective viscosity
$\eta_{eff}=\tau_R G_0$ for samples at $T=20^\circ \rm{C}$ and
with the following values of $\phi$ and $\alpha$, respectively:
12\%, 0.8 (squares); 16\%, 0.3 (circles); 16\%, 0.4 (triangles);
16\%, 0.8 (diamonds); 16\%, 1.2 (crosses); 18\%, 0.8 (plus). The
straight line is a power law fit to all the data yielding an
exponent $0.85\pm0.04$.} \label{FIG:eff_visco}
\end{figure}

To test the robustness of the concept of effective viscosity, we
follow by linear rheology the aging of samples of various
composition, and thus different elasticity (both $\phi$ and
$\alpha$ were varied leading to $60 \leq G_0 \leq 700$ Pa). Each
sample is quenched in the solid phase and aged without any
perturbation during a time $t_w$ before being submitted to a step
strain allowing the measurement of $\tau_R$, as described
previously. For all samples, we find that $\tau_{R} \sim t_{w}^m$
with an average exponent $\overline{m}=0.87 \pm 0.09$,
consistently with our previous experiments \cite{Ramos2001} and
similarly to what observed for other soft glassy materials
\cite{int_stresses,Cloitre2000, Derec2003}. Remarkably, all data
collapse onto a single master curve when plotting the age
dependence of the effective viscosity $\eta_{eff}=\tau_RG_0$, as
shown in fig.~\ref{FIG:eff_visco}. This scaling demonstrates that
not only does the effective viscosity characterize the slow
dynamics at a given age, but it also accounts for the aging of
samples with different elasticity in a unified way.

\begin{figure}
\includegraphics{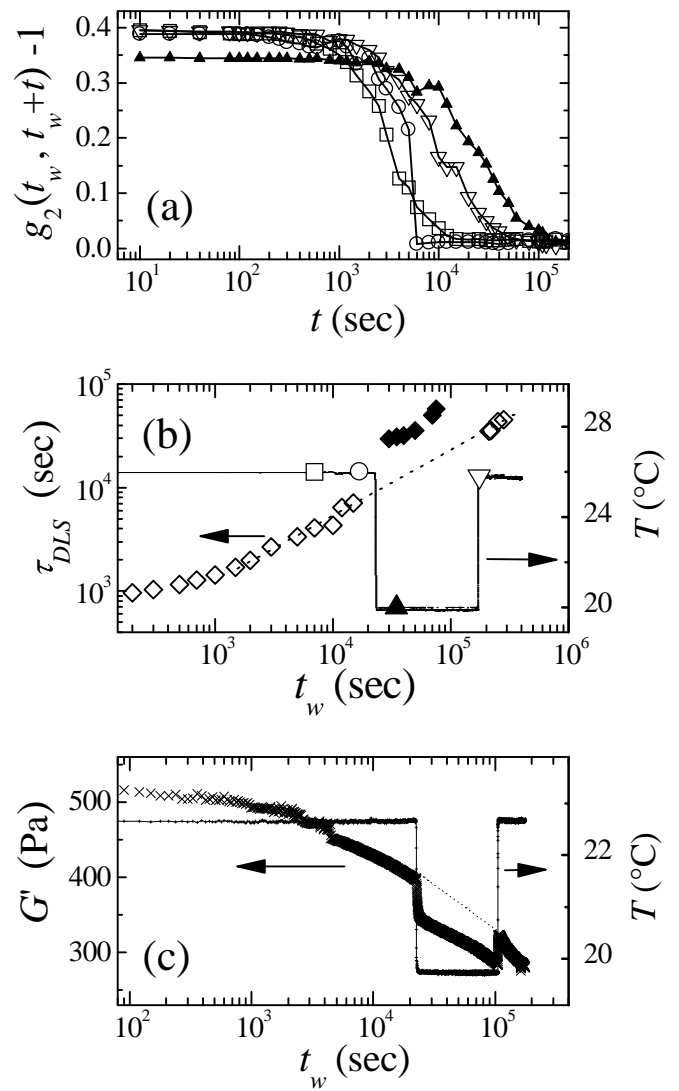}
\caption{(a) Intensity correlation functions taken at age $t_w =$
$7000$ sec (squares), $17000$ sec (circles), $35000$ sec (up
triangles) and $175000$ sec (down triangles) for a sample (with
$\phi=16\%$, and $\alpha=0.8$) submitted to the temperature
history shown in (b). (b) (left axis) Age dependence of the
characteristic time extracted from the correlation functions (the
dashed line is a guide for the eye). (right axis) The solid line
shows the temperature of the sample and the symbols indicate the
times at which the correlation functions shown in (a) are
measured. (c) Age dependence of the storage modulus measured at
frequency $\nu=1$ Hz, for a sample submitted to the temperature
history shown as a solid line. The dashed line is a guide for the
eye.} \label{FIG:3}
\end{figure}

The aging behavior of many glassy systems is deeply affected by a
temperature \cite{T_perturbation} or a mechanical
\cite{ViasnoffPRL2002,Mechanical_perturbation} perturbation,
leading to surprising effects such as the memory effect and
rejuvenation. The question naturally arises whether the effective
viscosity introduced here for the MLV phase is sensitive to a
perturbation of the elastic modulus during the aging. To address
this issue, we study the time evolution of the dynamics of a
sample quenched from $T=4$ to $T=26^\circ \rm{C}$ and afterwards
submitted to a square wave temperature perturbation, from $26$
down to $20$ and back to $26^ \circ \rm{C}$, resulting in a
variation of the elastic modulus of $20 \%$. We follow the aging
by MDLS: the multispeckle technique allows time-resolved
information on the dynamics to be obtained, thus probing both the
aging behavior and the instantaneous response of the soft glass
configuration to the change in temperature. Figure \ref{FIG:3}a
shows $g_2(t_w,t_w+t)-1$, the two-time intensity correlation
function measured at a scattering vector $q = 13.1 \, \mu{\rm
m}^{-1}$ before, during, and after the temperature perturbation.
Simultaneously to the temperature jump, the correlation function
drops abruptly to $0$ (circles in fig. \ref{FIG:3}a), thus
revealing that the temperature perturbation not only modifies
$G_0$, but it also significantly affects the local glass
configuration. In fact, the complete loss of correlation is
indicative of rearrangements on a length scale larger than $2 \pi
/q \simeq 0.5 \, \mu \rm{m}$, a significant fraction of the MLV
size \cite{Note_decorrelation}. The characteristic time of the
decay of $g_2-1$, $\tau_{DLS}$ \cite{Note_tau}, is plotted in
fig.~\ref{FIG:3}b as a function of age, for the same sample as in
a). Before the perturbation, $\tau_{DLS}$ exhibits a power law
growth with sample age, as observed in previous work
\cite{Ramos2001}. When the sample is cooled at $T = 20^{\circ}
\rm{C}$, $\tau_{DLS}$ suddenly increases by more than a factor of
two; however, note that the growth of $\tau_{DLS}$ follows a trend
similar to that before the temperature jump. Values of
$\tau_{DLS}$ higher at $T=20^{\circ} \rm{C}$ than at $T=26^{\circ}
\rm{C}$ are in qualitative agreement with the trend observed for
the relaxation time measured by rheology, since the lower $T$ the
smaller $G_0$ and hence the larger $\tau_{R}$. However, we note
that the variation of $\tau_{DLS}$ is much larger than that of
$\tau_{R}$ (for the former is $130 \%$, while for the latter is
$20 \%$). Surprisingly, when $T$ is increased back to its initial
value, $\tau_{DLS}$ drops abruptly and recovers the power law
evolution it would have had if the temperature was never changed.
Similar results have been obtained for a positive square wave
temperature perturbation. Light scattering data thus indicate
that, in spite of the significant change in the local
configuration and in the relaxation time produced by the $T$ jump,
the evolution of the dynamics is not affected by the perturbation
but rather follows an ``intrinsic aging'' behavior.

\begin{figure}
\includegraphics{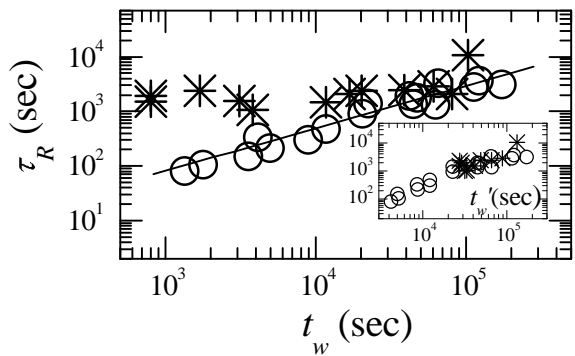}
\caption{Characteristic relaxation time measured by rheology for a
sample (with $\phi=16\%$ and $\alpha=0.8$), (circles) rapidly,
  (stars) slowly,
  brought from $T=4^{\circ}
\rm{C}$ to $T=20^{\circ} \rm{C}$. In the main plot
  the age $t_w$ of the sample is defined as usual as the
  time elapsed since $T$ has reached its final value,
  while in the inset $t'_w$ is defined as the time elapsed since
  the sample is solid.}
\label{FIG:4}
\end{figure}

It is tempting to associate this intrinsic aging to a steady
growth of $\eta_{eff}$ similar to that reported in
fig.~\ref{FIG:eff_visco} for the undisturbed soft glasses.
Unfortunately, time-resolved stress relaxation experiments that
would directly confirm a continuous increase of $\eta_{eff}$ for
the samples submitted to a temperature perturbation are not
available. Instead, we note that if the evolution of the dynamics
does not depend on the elastic history of the sample, as suggested
by the MDLS experiments, the rate at which the sample is quenched
from the fluid to the solid phase should not modify the aging. In
fact, a slow and continuous increase of $T$ may be viewed as a
series of small $T$ increments. In fig.~\ref{FIG:4} we compare the
characteristic time measured by rheology for samples brought from
$4$ to $20 ^{\circ} \rm{C}$ either rapidly (``fast quench'',
heating rate $r\simeq 0.64 ^{\circ} \rm{C/min}$) or slowly (``slow
quench'', $r=0.03 ^{\circ} \rm{C/min}$). If $t_w$ is defined as
the time elapsed after reaching the final temperature (main plot),
a marked difference exists between the aging of samples quenched
at different rates, as observed for spin glasses
\cite{T_perturbation}. By contrast, the data for both fast and
slow quench rates collapse onto a single curve (inset of
fig.~\ref{FIG:4}) when defining the age as the time elapsed since
the materials is a solid (defined by $G' \geq G''$), as expected
if the effective viscosity does not depend on the elasticity of
the sample or its elastic history, but only on the time spent in
the glassy phase.

To further support the intrinsic aging scenario and to gain
insight on the physical mechanism responsible for the growth of
$\eta_{eff}$, we follow the time evolution of $G'(\nu = 1 \, {\rm
Hz})$ for a sample submitted to a temperature perturbation similar
to that of the MDLS experiments. At constant temperature, $G'$
slowly but continuously decreases with time, as shown in
fig.~\ref{FIG:3}c. This behavior suggests that the storage modulus
can be written as the sum of two contributions: the elastic
modulus at equilibrium, $G_{eq}$, and the internal stress, built
up at the fast transition from the fluid to the solid phase and
whose slow relaxation is responsible for the aging dynamics. When
the temperature is varied from $23$ to $20^{\circ} \rm{C}$, the
storage modulus drops by 10\%, a change in remarkable agreement
with the difference of $G'$ for samples directly quenched to these
two temperatures. In analogy with the results of the MDLS
experiments, when $T$ is brought back to its initial value $G'$,
resumes the same time evolution it would have had without any
temperature perturbation. These results provide additional
evidence that the aging of the MLV phase is described by the
evolution of an effective viscosity that only depends on $t_w$.

In order to rationalize our experimental findings, we present a
simple model that links the effective viscosity to the internal
stress, $\sigma_{int}$.  In this picture $\sigma_{int}$ stems from
the elastic deformation of the MLVs with respect to a spherical
shape, due to the random growth of closely packed MLVs at the
fluid-to-solid transition. With time, the elastic deformation is
slowly reduced through structural relaxations that are at the
origin of the slow dynamics, leading to the aging behavior. We
consider that the stress relaxation measured by rheology is due to
rearrangements of regions of size $L$ (presumably containing
several MLVs) that are submitted to a driving force, due to
$\sigma_{int}$, and a viscous drag force. The driving force and
the viscous drag scale respectively as $\sigma_{int} L^2$ and
$\eta_{micro}  L V$. Here, $\eta_{micro}$ is a microscopic
viscosity related to the solvent viscosity and the MLVs' volume
fraction, and $V$ is the local velocity of the rearranging zone.
By balancing the driving force and the viscous drag, one obtains
$\sigma_{int} \sim \eta_{micro} V / L$. Because both the elastic
modulus and $\sigma_{int}$ have as a common microscopic origin the
elastic energy due to the deformation of the MLVs, one expects
$\sigma_{int}$ to be, at all time, proportional to the equilibrium
elastic modulus of the material: $\sigma_{int}(t_w) = G_{eq} / \xi
(t_w)$, with $\xi (t_w)$ a dimensionless proportionality factor
whose growth with $t_w$ describes the aging. Writing as suggested
previously $G_{0}=G_{eq}+ \sigma_{int}$, one gets
$\sigma_{int}=G_{0} / (1+\xi) \sim \eta_{micro} V / L$.
Experimentally, the characteristic time of the stress relaxation
is related to the time needed for a region of size $L$ to move
over a distance equal to its size, $\tau_R \sim L/V$. From
$\eta_{eff}=G_0 \tau_R$, one finally obtains $\eta_{eff}(t_w) \sim
\eta_{micro}(1+\xi(t_w))$. In agreement with experiments, this
scaling shows that the effective viscosity is independent of the
elastic modulus of the material and that the relaxation of the
internal stress leads to an increase of $\eta_{eff}$ with sample
age. The intrinsic aging reported above suggests that neither an
isolated temperature jump (fig.~\ref{FIG:3}) nor a series of very
small $T$ increments (fig.~\ref{FIG:4}) are able to relax
significantly the MLVs deformation, although they change the local
configuration, as indicated by the drop of $g_2-1$. As a
consequence, they leave essentially unchanged $\xi$ and hence
$\eta_{eff}$. Increasing the amplitude or the number of $T$ jumps
may eventually suppress the intrinsic aging: experiments are in
progress to test this conjecture. Finally, we note that the
elasticity of the MLV phase is formally identical to that of
concentrated emulsions \cite{equilibriumonions}: our findings
should therefore be directly generalizable to the latter. Whether
the same conclusions may be drawn also for other concentrated soft
systems, e.g. deformable colloidal particles, remains an open
issue.

We thank A. Duri and S. Mazoyer for preliminary experiments, and
L. Berthier, E. Pitard and D. A. Weitz for fruitful discussions.
Financial support from the French Minist\`{e}re de la Recherche
(JC2076) and the EC Network of Excellence 'SoftComp' is gratefully
acknowledged.



\begin{references}


\bibitem{Cloitre2003} M. Cloitre \textit{et al.},
Phys. Rev. Lett. \textbf{90}, 068303 (2003).

\bibitem{KrallPRL1998}  A. H. Krall and D. A. Weitz, Phys. Rev. Lett.
\textbf{80}, 778 (1998).

\bibitem{ccemulsions} T. G. Mason and D. A. Weitz, Phys. Rev.
Lett. \textbf{74}, 1250 (1995).

\bibitem{Onions_CRPP} J. Leng, PhD thesis, Universit\'{e} Bordeaux 1
(1999).

\bibitem{Faraday}  L. Cipelletti \textit{et al.}, Faraday Discuss.
\textbf{123}, 237 (2003).

\bibitem{int_stresses}A. Knaebel \textit{et al.}, Europhys.
Lett. \textbf{52}, 73 (2000).

\bibitem{paper Munch PRE} M. Bellour \textit{et al.}, Phys. Rev. E
\textbf{67}, 031405 (2003).

\bibitem{Estelle} J.-P. Bouchaud and E. Pitard, Eur. Phys. J.
E \textbf{6}, 231 (2001).

\bibitem{Cloitre2000}  M. Cloitre, R. Borrega and L. Leibler, Phys. Rev.
Lett. \textbf{85}, 4819 (2000).

\bibitem{Bonn2002} D. Bonn \textit{et al.}, Phys. Rev. Lett. \textbf{89}, 015701
(2002).

\bibitem{Derec2003} C. Derec \textit{et al.}, Phys. Rev. E
\textbf{67}, 061403 (2003).

\bibitem{ViasnoffPRL2002} V. Viasnoff and F. Lequeux, Phys.
Rev. Lett. \textbf{89}, 065701 (2002).

\bibitem{Ramos2001} L. Ramos and L. Cipelletti, Phys. Rev. Lett.
\textbf{87}, 245503 (2001).

\bibitem{Ligoure} F. Castro-Roman, G. Porte and C. Ligoure, Phys. Rev.
Lett. \textbf{82}, 109 (1999).

\bibitem{equilibriumonions} L. Ramos \textit{et al.}, Europhys. Lett.
\textbf{66}, 888 (2004).

\bibitem{Note_uncertainties} Typical uncertainties on $\tau_R$ and
$p$ are less than $5\%$.

\bibitem{T_perturbation} K. Jonason \textit{et al.}, Phys. Rev.
Lett. \textbf{81}, 3243 (1998).

\bibitem{Mechanical_perturbation} F. Ozon \textit{et al.}, Phys. Rev.
E \textbf{68}, 032401 (2003).

\bibitem{Note_decorrelation} We have checked that the drops of $g_2-1$ are
not due to a local rigid shift of the sample non to refractive
effects, but rather to irreversible rearrangements.

\bibitem{Note_tau} We define $\tau_{DLS}$ as the time needed for
the correlation function to decrease by half of its initial value.


\end{references}
\end{document}